# Role of structure of C-terminated *4H*-SiC(000$\bar{1}$) surface in growth of graphene layers - transmission electron microscopy and density functional theory studies

J. Borysiuk,[1,2,a)] J. Sołtys,[3] R. Bożek,[2] J. Piechota,[3] S. Krukowski,[3,4] W. Strupiński,[5] J. M. Baranowski,[2,5] and R. Stępniewski[2]

[1]*Institute of Physics, Polish Academy of Sciences, Al. Lotników 32/46, 02-668 Warsaw, Poland*
[2]*Faculty of Physics, University of Warsaw, Hoża 69, 00-681 Warsaw, Poland*
[3]*Interdisciplinary Centre for Materials Modelling, University of Warsaw, Pawińskiego 5a, 02-106 Warsaw, Poland*
[4]*Institute of High Pressure Physics, Polish Academy of Sciences, Sokołowska 29/37, 01-142 Warsaw, Poland*
[5]*Institute of Electronic Materials Technology, Wólczyńska 133, 01-919 Warsaw, Poland*

Principal structural defects in graphene layers, synthesized on a carbon-terminated face, i.e. the SiC(000$\bar{1}$) face of a *4H*-SiC substrate, are investigated using microscopic methods. Results of high-resolution transmission electron microscopy (HRTEM) reveal their atomic arrangement. Mechanism of such defects creation, directly related to the underlying crystallographic structure of the SiC substrate, is elucidated. The connection between the *4H*-SiC(000$\bar{1}$) surface morphology, including the presence of the single atomic steps, the sequences of atomic steps, and also the macrosteps, and the corresponding emergence of planar defective structure (discontinuities of carbon layers and wrinkles) is revealed. It is shown that disappearance of the multistep island leads to the creation of wrinkles in the graphene layers. The density functional theory (DFT) calculation results show that the diffusion of both silicon and carbon atoms is possible on a Si-terminated SiC surface at a high temperature close to 1600°C. The creation of buffer layer at the Si-terminated surface effectively blocks horizontal diffusion, preventing growth of thick graphene layer at this face. At the carbon terminated SiC surface, the buffer layer is absent leaving space for effective horizontal diffusion of both silicon and carbon atoms. DFT results show that excess carbon atoms converts a topmost carbon layer to sp$^2$ bonded configuration, liberating Si atoms in

---

[a)]Electronic mail: jolanta.borysiuk@fuw.edu.pl

barrierless process. The silicon atoms escape through the channels created at the bending layers defects, while the carbon atoms are incorporated into the growing graphene layers. These results explain growth of thick graphene underneath existing graphene cover and also the creation of the principal defects at the C-terminated SiC(0001) surface.

# I. INTRODUCTION

For the last decade, graphene has been a hot research topic due to its unusual electronic properties, very attractive for potential applications in future sensing and high speed electronic devices.[1] Since its discovery, a single atomic layer graphene, characterized by the $sp^2$ bonding of the carbon atoms, distributed on a hexagonal honeycomb lattice, has been denoted as graphene. The single carbon layer shows a number of unusual electronic properties, arising from a Dirac-type dispersion relation, confirmed directly by an anomalous magnetic field dependence of Integer Quantum Hall Effect.[1-7] The unusual electronic transport properties, related to π-bonds, formed by p-electrons of the carbon atoms, attracted an immense interest of many researchers in graphene.

In the past years, the notion of graphene was extended from a single layer of carbon to bilayer or many layers of carbon atoms. Such multilayer graphene was synthesized on various substrates: gold, nickel, platinum and ruthenium.[8-12] However, a natural, the most convenient way to obtain a single layer or a few layers of graphene, is to use carbon containing electrically insulating crystals, such as diamond and silicon carbide. Diamond is technically difficult to process but silicon carbide thermal decomposition that leads to a loss of more volatile silicon is natural candidate for the graphene technology. It is known that vacuum annealing at a high temperature of *6H*-SiC or *4H*-SiC crystals leads to the synthesis of the carbon layer in a graphitic form at the surface.[13-15] Both polar surfaces, i.e. Si- and C-terminated ones, could be used but their potential depends on the properties of the synthesized graphene.[16-19]

Graphene synthesized at a C-terminated SiC surface seems to be relatively weakly attached to the underlying surface. No buffer layer, i.e. covalently bound carbon layer, was detected and the first graphene layer is located at the distance of about 3.2Å from the SiC surface, i.e. too far to form covalent bonding between carbon atoms.[20] Low energy electron diffraction (LEED)[21-23] and scanning tunneling microscopy (STM)[24-26] results indicate that a significant degree of rotational disorder in the graphene films. Rotational stacking faults in the graphene layers, which give rise to a moiré pattern, were also observed in TEM.[27]

Systematic investigations of the growth of graphene layers in both orientations have been undertaken quite recently and have brought a limited insight into the processes. Studies over a growth on a Si-terminated face have been far more advanced. It is recognized that a graphene growth depends on the temperature, gas pressure and also on the morphology of the SiC(0001) surface.[28-33] In standard processes, vicinal SiC surfaces are gas etched in the $H_2$ atmosphere. Depending on the duration of this stage, the partial erosion of the material at the steps occurs which may lead to the formation of macrosteps[28,29] or a significant modification of the step shapes.[30]

Typically, prepared SiC samples are annealed in a neutral atmosphere at a very high temperature which leads to graphitization that begins at the steps. The graphitization process proceeds by evaporation of silicon atoms, first at the steps where there atoms are more weakly bonded.[28-34] The details of the process are still disputed, depending on the initial structure of the SiC surface. For vicinal surfaces, it was proposed that surface diffusion is negligible so that three SiC layers constitute a single graphene layer.[29,30] In some cases, the disintegration of the three steps leads to the formation of trenches, at the portion of the terraces, close to the steps.[33] In addition, the creation of surface pits is observed.[29] Most likely, the creation of the pits is related to dislocation or other extended defects in the SiC substrate. At a higher temperature, step instability leads to the formation of finger-like patterns.[31] The graphene formation process is two-dimensional.

For SiC surfaces transformed to macrosteps, the graphitization scenario is different.[28,32] It was proposed that the anisotropic process of the growth of graphene layers is supported by C adatom diffusion.[28] Graphene flakes start to grow at a sidewall of the macrosteps which then continue to extend over a part of the terraces.[32] It is stipulated that the first graphene layer is created by evaporation through the buffer layer and the subsequent layers emerge due to direct penetration of silicon atoms either by vacancy sites[34] or grain boundaries.[35] In order to reach these sites, the liberated silicon atoms diffuse under the conditions in a plane parallel to the carbon atoms. For thicker graphene, the silicon outdiffusion slows down, leading to effective growth termination after the creation of 4 to 5 carbon layers.

In a slightly different version, supported by TEM data, the emerging graphene layers are also created at the side of macrosteps.[33] It is postulated, however, that these layers are anchored at the lower terraces by SiC surface defects, creation of the structures. The structures have the carbon layers shaped perpendicularly to the SiC surface at the anchoring points. The layers extend over the surface of the upper terrace indefinitely, ultimately to the terrace termination. The new carbon layer originates underneath of the previously created carbon layer, at the side of the macrosteps.[33] This model does not allow for disintegration of the terrace or formation of trenches.

The synthesis of the graphene layers, at a C-terminated surface is investigated in the present paper. It is well known that the number of graphene layers could be very high, reaching 40 layers, obtained in a single process. In addition, graphene layers are relatively weakly coupled to the SiC surface, located at a distance of about 3.0 - 3.2 Å.[20,36] Thus, this space creates a channel for effective horizontal diffusion of liberated Si atoms. This picture is supplemented by the detailed TEM analysis of graphene layers synthesized on this face. These data are confronted with the results of DFT simulations of the energy barriers for diffusion of Si and C atoms on a C-terminated SiC surface. A proposed hypothetical scenario of the graphene synthesis is built using these two sets of data.

## II. DFT CALCULATION METHOD

An *ab initio* simulation of the energy barrier for the diffusional motion of carbon and silicon atoms at polar {0001}SiC surfaces were using a commercially available Vienna *ab initio* simulation package (VASP), based on a plane wave basis set.[37-40] The projector augmented wave (PAW) approach[41] was used in its variant available in the VASP package.[40] For the exchange-correlation functional, the local spin density approximation (LSDA) was applied. The plane wave cut-off energy was set to 500 eV. The Monkhorst-Pack k-point mesh was set to 7×7×1. The *4H*-SiC{0001} superlattice was constructed using 8 bilayers of Si-C which was sufficient to avoid a quantum overlap of the termination and real surface quantum states.[42] Two top SiC layers were relaxed using the conjugate gradient algorithm.

## III. EXPERIMENTAL

The graphene layers were grown on the carbon terminated surface of the *4H*-SiC ($000\bar{1}$) substrate in an Epigress VP508 SiC hot-wall chemical vapor deposition (CVD) reactor, heated by an RF generator.[43] Initially, the SiC substrate was etched in the atmosphere of mixed hydrogen and propane ($H_2$-$C_3H_8$) at a temperature close to 1600ºC. Subsequently, the substrate was annealed in the argon atmosphere for about 20 minutes at a gas pressure of 100 mbar and a temperature of 1600ºC. The annealing led to evaporation of the silicon atoms and creation of the graphene multilayers by less volatile carbon atoms.

High-resolution HRTEM observations of the graphene layers were performed using JEOL JEM 3010 transmission electron microscope operating at 300 kV. Cross-sectional TEM specimens were prepared by a standard method, based on mechanically prethinning of the samples followed by an Ar ion milling procedure.

Atomic Force Microscopy (AFM) images were recorded using the tapping mode, in air at ambient pressure using a Nanoscope IIIa.

## IV. RESULTS

### A. Diffusivity of carbon and silicon diffusion at SiC surfaces – DFT study

The energy surface for the silicon and carbon atoms at the Si-terminated face is presented in Fig. 1. From the profiles, it follows that for both adatoms, i.e. Si and C, the energy maximum corresponds to the position "on top" above the topmost Si atoms. As it is shown, the two energy minima correspond to the two types of the positions above the carbon atoms. The energy difference between the global maximum and minimum is equal to 1.8 eV. It is worth noting that these two minima have almost identical energies the difference of their energy is mere 0.13 eV, about 7% of the total energy span. That indicates the nature of the bonding of the Si adatom, attached to the two neighboring Si atoms in the bridge position. It is also remarkable that the energy barrier obtained from energy profiles along the path connecting the energy minima is bout 0.40 eV. This is related to the fact that the overlap to the neighboring silicon atoms is not much affected by the displacement, thank to relatively close distance, due to relatively short interatomic distances in the SiC lattice as compared to a Si crystal. Therefore from this diagram it follows that relatively deep valleys exist in the energy surface, creating effective surface diffusion channels for Si adatoms. At the temperature of typical SiC sublimation processes, used for the synthesis of graphene on the Si-terminated SiC surface, slightly exceeding 1100°C, the surface diffusion process is fairly effective allowing a transfer of Si adatoms across the terraces even these of a relatively large size. Therefore, the observed preferential growth of graphene layers at SiC steps is not related to the difference in Si concentration at the terraces but to atomic disintegration kinetics at the edges.

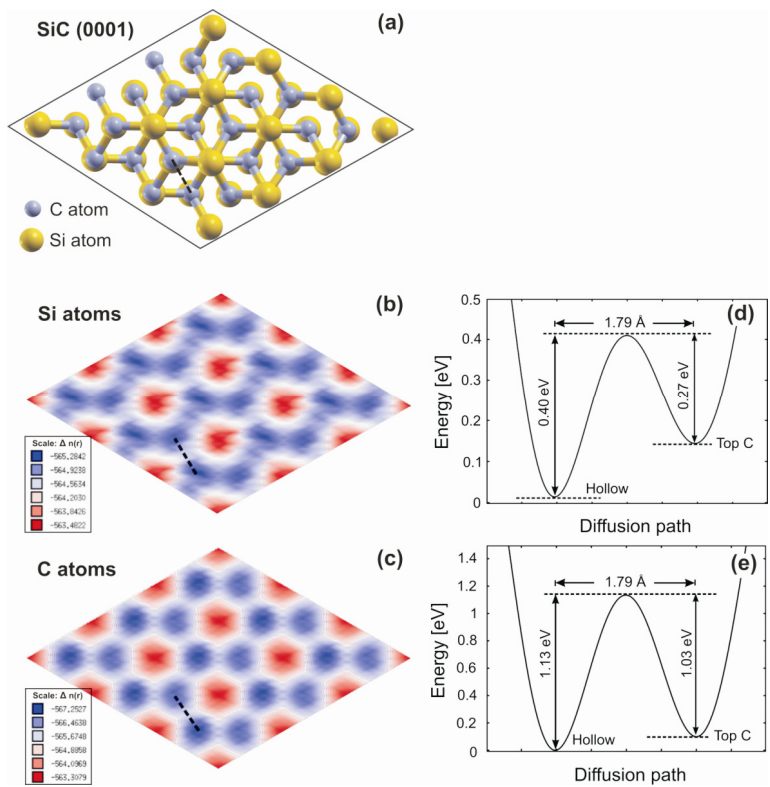

**Fig. 1.** Atomic structure of silicon carbide Si-terminated surface and the energy surface of the adsorbed Si and C atoms derived from DFT results: (a) the atomic structure of SiC surface, (b) the energy surface of the silicon atom, (c) the energy surface of the carbon atom, at the (0001)SiC surface. On the right hand side (d,e), the energy profiles along the black line correspond to the distances shown in the diagrams.

It is worth noting that energy profiles for C adatoms, at the (0001)SiC surfaces are qualitatively different. As in the previous case, C adatoms have their energy maximum in the positions "on top" of the topmost silicon atoms. This is somewhat unexpected as this position corresponds to the site in the SiC surface, saturating a broken Si bond. It turns out, however, that the bridge position is more favorable energetically, bonding a C adatom with the two neighboring topmost Si atoms that decreases the energy by impressive 3.9 eV. Note that this is independent of the position of the C atom beneath as these two minima differ by mere 0.1 eV. Similarly to the above case of Si adatoms, the energy valleys for C diffusion exist at the (0001)SiC surface. The stretching of the C-Si bond leads to a much higher energy expense than before, therefore the energy barrier for diffusion of carbon is 1.13 eV. This indicates that the diffusion of carbon is much less effective which is in accordance with the STM observations of the preferential growth of graphene at the SiC terrace edges.

The energy profiles, obtained for a C-terminated SiC surface reveal a different landscape. As shown in Fig. 2 the energy difference between the maximum, being in the "on top" of the topmost C atoms and the global minimum is quite substantial, 2.45 eV. Additionally, the two minima are markedly different, the energy for the "on top" of Si is lower by 0.45 eV than the hollow position. The energy minima do not create low energy corridors but they create well defined site locations which are surrounded by substantial energy barriers, 0.96 eV. Thus the diffusion of silicon adatoms is relatively difficult, yet at a relatively high temperature, typically used for graphene synthesis of C-face, close to 1600°C, the barrier could be surmounted and Si diffusion could contribute to the overall dynamics of SiC disintegration and the creation of epitaxial graphene.

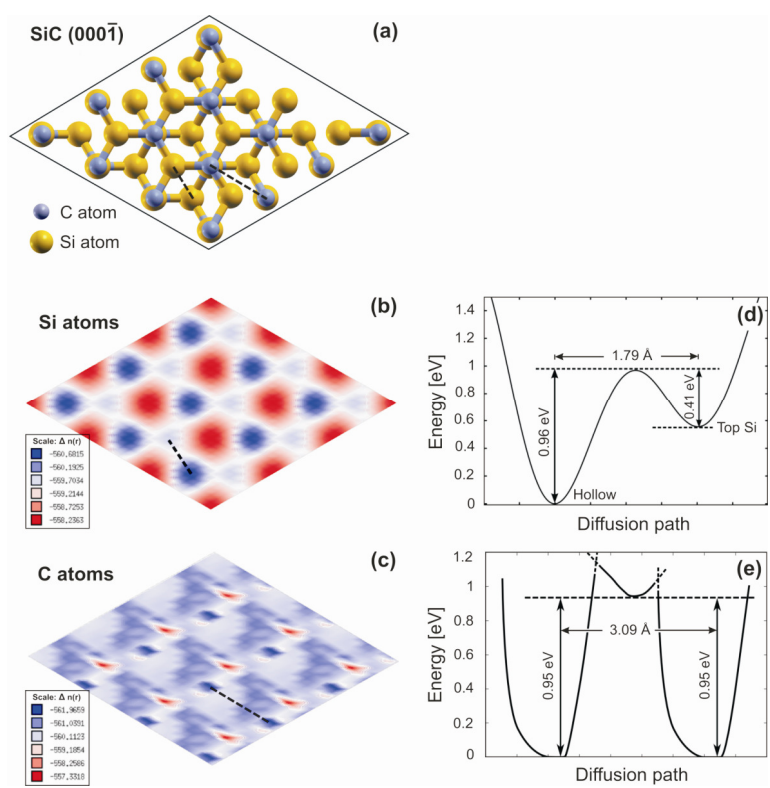

**Fig. 2.** Atomic structure of silicon carbide C-terminated surface and the energy surface of the adsorbed Si and C atoms derived from DFT results: (a) the atomic structure of an SiC surface, (b) the energy surface of the silicon atom, (c) the energy surface of the carbon atom, at the $(000\bar{1})$SiC surface. On the right hand side (d,e), the energy profiles along the black line correspond to the distances shown in the diagrams.

Another completely different picture emerges for the energy landscape of a C atom at a C-terminated SiC surface. Carbon adatoms are strongly bound in the "on top" position, above the topmost carbon atoms. The bonding is very specific as it is strongly localized, having the energy much lower than in any other location. Thus there is a steep energy minimum and a similar steep maximum in the hollow position of a C adatom strongly localized in the positions "on top" of C and Si atoms. The energy difference between these two locations reaches about 4.6 eV. Apart from this, the dominant fraction of the surface is energetically flat, about 1.0 eV above the minima. As it is also shown, the energy path shows non-analytic behavior, being composed of the two different energy surfaces.

The nature of the unexpected behavior is elucidated by the plot of atomic arrangements of the SiC surfaces, presented in Fig. 3. As it is shown there, the presence of a Si adatom at a Si-terminated surface weakly affects the atomic structure of the surface. A completely different scenario was obtained for a C adatom at a C-terminated surface. The carbon atom is strongly bound to the neighboring C atoms, removing a Si atom from the lattice to the adsorbed position. Thus a carbon adatom is replaced by a Si adatom and a portion of the flat C plane, vividly resembling graphene. The position of these C atoms corresponds to the $sp^2$ bonding configuration, indicating on the catalytic influence of carbon adatoms. Thus the nucleation of the graphene layer in the presence of carbon adatoms occurs at SiC terraces, not necessarily at SiC steps. This indicates a possibility of creating a large number of independently growing graphene flakes at the $(000\bar{1})$ SiC surface, which coalesce incoherently gives rise to grain boundaries and highly dislocated graphene planes. At large strains, the carbon layers are buckled giving rise to the creation of wrinkles.

It has to be added that the motion of a C atom, located in the reconstructed needs to overcome a much higher barrier, i.e. 1 eV shown in Fig. 3. The middle part of the diagram corresponds to the repositioning Si atom back in the lattice, which is unlikely as the liberated Si adatom moves away. The analytic continuation of the energy curve from the localized position is very steep indication of very high energy needed to remove carbon from the $sp^2$ bonded lattice. Thus the presented transformation corresponds to irreversible trapping of carbon in the graphene lattice.

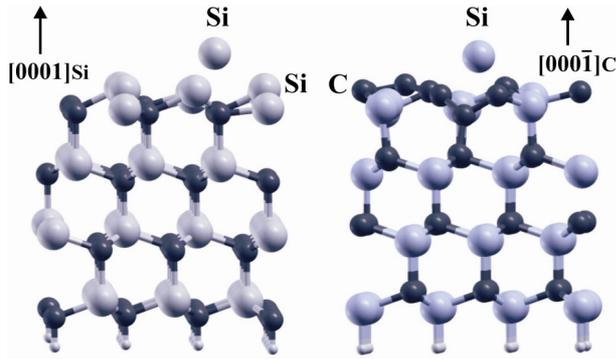

**Fig. 3.** Side view of silicon carbide Si- and C-terminated surfaces, with similar adatoms in the equilibrium positions. The grey and black balls correspond to Si and C atoms respectively.

It is well recognized that one of the critical steps of the graphene synthesis by silicon removal from SiC is sublimation of silicon atoms. It is therefore essential to determine the energy barrier for direct detachment of silicon adatoms from the two configurations presented in Fig. 3. The energy barrier was determined by DFT calculations from the energy difference between these configurations and those that Si atoms are far away from the surface. It turned out that the energy barrier is very high, equal to 6.13 eV and 6.15 eV for these two configurations, respectively. Thus direct evaporation of silicon atoms is not an effective channel for Si sublimation. The effective Si evaporation channel involves a more complicated molecular scenario, for both cases alike.

**B. Structure of graphene layers grown over vicinal *4H*-SiC surfaces**

A typical multilayer graphene, grown on the SiC($000\bar{1}$) surface, measured by an optical microscope, is presented in Fig. 4(a). It is shown that the graphene layers are corrugated, with number of macrosteps, easily visible even at low magnification. These steps are not affected by the crystal edge and they are related to the step structure developed during the etching procedure. At higher magnification a dense network of smaller defects is visible in AFM image - Fig. 4(b). The defects - wrinkles, about 2 nm high, are mostly created in the neighborhood of the steps. Structure of the defects in multilayer graphene is very strongly dependent on the initial structure of an annealed surface.

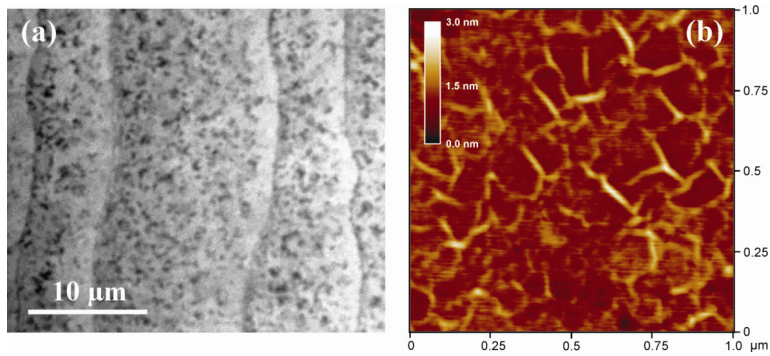

**Fig. 4.** Morphology of the SiC(000$\bar{1}$) surface containing graphene layers observed by optical microscope with Nomarski contrast (a) and AFM image of magnified part of surface (b).

Relatively perfect regions exist in the areas between surface steps. Such regions, showing a flat multilayer graphene structure are demonstrated on TEM image in Fig. 5. Annealing at a temperature of T=1600ºC for t=20min leads to growth of about 15 graphene layers [Fig. 5(a)]. It is shown that the defect-free substrate could be transformed into a multilayer graphene with a large number of graphene layers without any structural defects. The HRTEM image of a few graphene layers deposited on the SiC substrate is shown in Fig. 5(b). From HRTEM image analysis it follows that the first carbon layer is located about 3.0±0.2 Å above the SiC surface. The next carbon layers are at the distance of 3.3±0.2 Å, typical for the interplanar spacing of hexagonal graphite $d_{0002}$=3.35 Å. These results are in agreement with previous observations performed for graphene on C-terminated face.[20,36]

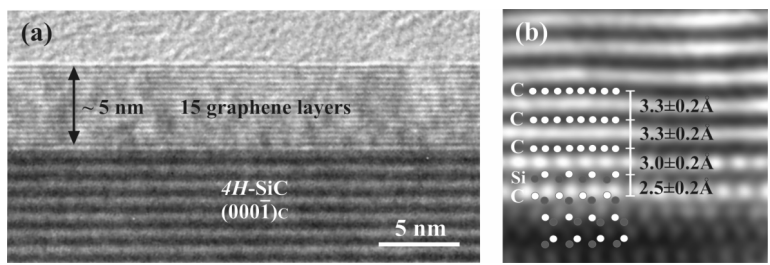

**Fig. 5.** Cross-sectional HRTEM images of graphene layers observed in the [11$\bar{2}$0] SiC orientation, showing the arrangement of 15 carbon layers (a) and filtered image of a few C layers associated with SiC substrate (b).

The flat graphene areas are surrounded by the defective regions which are related to the SiC morphology, developed during the etching process. These structures originate at the macrosteps, resulting from coalescence of many atomic steps, present at initially vicinal SiC surfaces. TEM structural and diffraction images, of graphene multilayer, about 6 nm thick, containing from 20 (upper) to 25 (lower terrace) layers, are presented in Fig. 6. A number of graphene layers at the upper and lower SiC terrace is different and new C layers are formed underneath the existing thick graphene cover. Formation of the new C layers, adjacent to the ~5nm high step is showed in Fig. 6(a). Magnified area in Fig. 6(c) [inset] presents transformation of the (0001)SiC atomic planes into the (0001)C planes. In the [$\bar{1}$100] SiC orientation, on defocus TEM images the disintegrated SiC areas are sometimes mapped with characteristic contrast. Every fourth plane in polytype *4H* of SiC is imaged with light contrast (dashed lines in the inset). This effect is most likely associated with a stress caused by an atomic rearrangements. The diffraction images obtained from the flat area of graphene and area with a step are shown in Fig. 6(b,d), respectively. There is a superposition of two diffraction patterns: first from the substrate structure of the [$\bar{1}$100] *4H*-SiC orientation and the second of the diffraction of the carbon (0002)C planes. Fig. 6(b) shows the diffraction pattern obtained from flat graphene on the terrace. The measured interplanar spacing between graphene layers $d_{0002}$= 3.4Å corresponds to the theoretical interplanar distance in the graphite structure $d_{0002}$= 3.35Å. The diffraction pattern from the area of the step shows streaks along the [0001] direction and the additional pair of maxima, reflected from the graphene, in the vicinity of the (0004)SiC spots [Fig. 6(d)]. The measured interplanar spacings of graphene are equal to $d_{0002\ C}$= 3.4Å and $d_{0002\ EG}$= 3.7Å. The splitted diffraction spots are located along the same direction, indicating a parallel orientation of the graphene atomic planes. The difference in the distance between graphene layers results probably from match of carbon planes to the SiC planes in early stage of the formation of carbon structure.

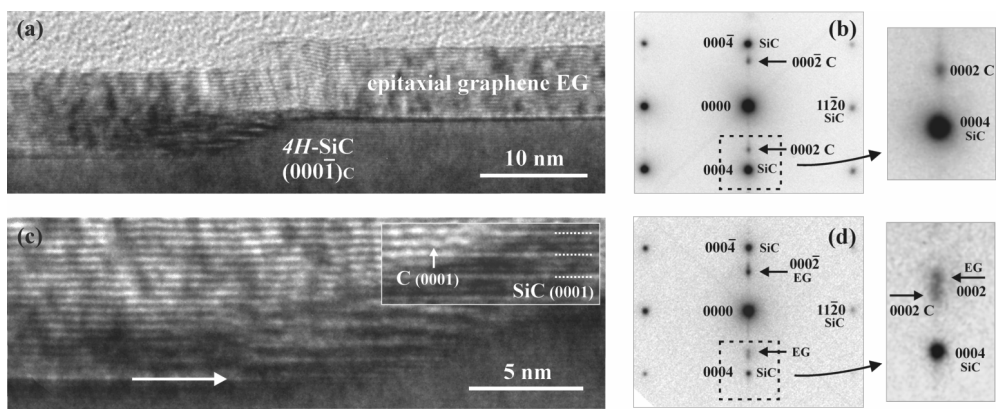

**Fig. 6.** Cross-sectional TEM image of the graphene structure on the SiC substrate containing macrostep. Contact of two thick carbon layers originated from two terraces (a) with diffraction pattern from a flat part of graphene (b). Magnified part of macrostep (marked by arrow) with inset showing details of transition of the planes from SiC to C (c) and diffraction pattern from area include the step (d). The magnified areas used in the analysis are shown beside the entire diffraction patterns. Diffraction patterns are negatives. Images are in the $[\bar{1}100]$ SiC substrate orientation.

The SiC surface disintegration at the macrosteps and islands is correlated with the presence of discontinuities of graphene structure (Fig. 7). As it is shown in Fig. 7(a) at the top, such left-over islands are related to creation of the graphene with bending layers. It is marked by dashed lines in Fig. 7(a). The bending of the higher located graphene multilayer results from blocking by the lower one. This arrangement is stable until the graphene layers from two levels of SiC surface are disconnected. In a different situation a flatting and connecting of the top carbon layers is observed.[33] Such flatting of graphene layers in the vicinity of small steps is visible in the bottom right corner in Fig. 7(b). A connection of graphene discontinuities leads to stacking of C layers parallel to the edge of SiC macrostep.

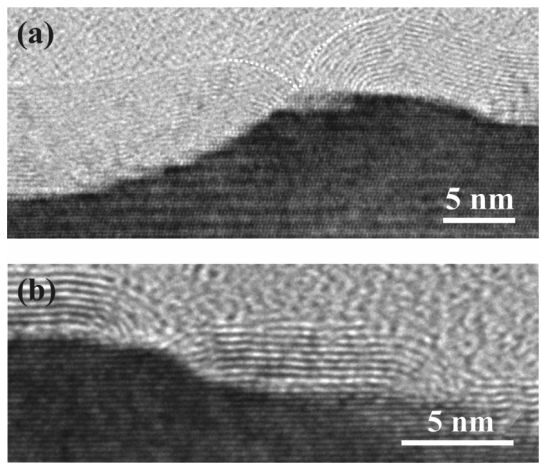

**Fig. 7.** TEM images of discontinuities and bendings of the graphene multilayers created near a macrostep (a) and at the small steps (b). Graphene with bending layers is marked by dashed lines.

Examples of various arrangements of graphene layers in respect to the SiC planes, observed in TEM, are shown in Fig. 8. In the case of a strong bending of the (0001)C planes almost perpendicular connections with the SiC stacking sequences are observed [Fig. 8(a)]. For macrosteps, the (0001)C planes are situated parallel to the edge of steps and for small steps the C planes are located along the (0001)SiC planes - Figs. 8(b) and 8(c), respectively. These three cases of the graphene arrangement are shown graphically in Fig. 8(d). For cases of (1) and (2) the new carbon atoms are added consecutively to the present (0001) graphene planes; it is most likely related to the existing covalent atomic bonds between substrate and graphene. The (3) case is proper to the creation of new graphene planes, where C atoms from SiC decomposition are quickly bonded with graphene. On the interface of the SiC substrate and graphene layers, the misfit lattice leads to a formation of dislocations in the (0001) basal plane.

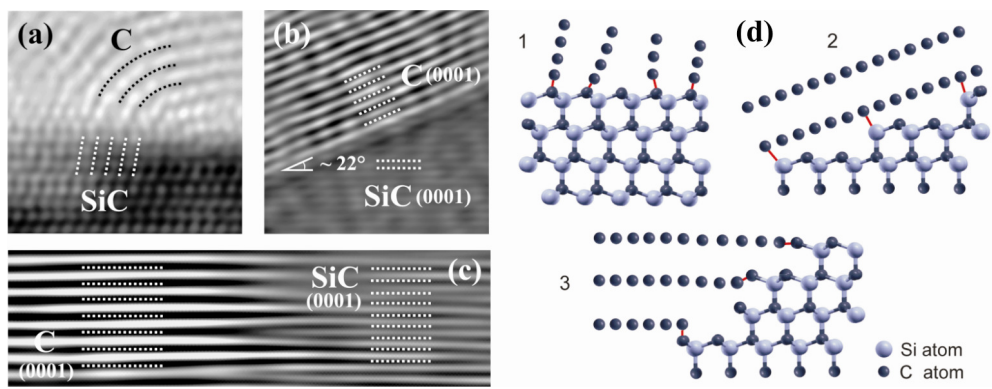

**Fig. 8.** Examples of various arrangements of anchoring of graphene layers to the SiC planes: (a) the (0001)C planes almost perpendicular to the surface, (b) the (0001)C planes are parallel to the edge of the macrostep, and (c) C planes are located along the (0001)SiC planes. Sketch showing the mutual position of the graphene layers and SiC planes (d). Dashed lines denote position of the planes.

Dissolution of the SiC islands causes creation of series of characteristic structures, as shown in Fig. 9. The bending of the graphene multilayers is related to the anchoring of C layers on the SiC steps and this structure terminates growth of carbon layers on the multistep SiC islands - Fig. 9(a,b). In the area between islands the graphene layers are rarely observed.

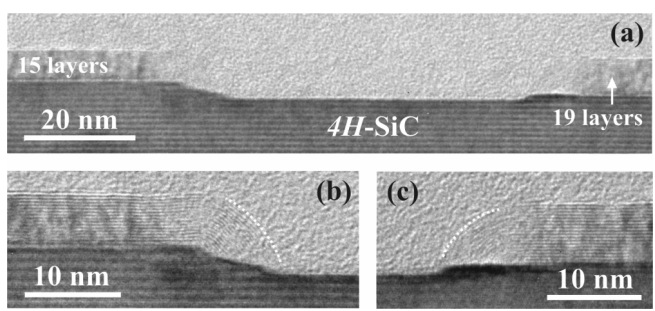

**Fig. 9.** TEM images of a pair of graphene multilayers heaving a different thickness: (a) a general view of the graphene structure on the SiC islands, (b) and (c) magnified parts of the carbon layers located at the edge of the two SiC islands.

A close relationship between disappearing multistep islands and the bending graphene structures on both side steps is demonstrated in Fig. 10. The terminating structure is created by ultimate disappearance of the SiC multistep island and a direct interaction between these bending structures. After coalescence, higher located graphene multilayers have an excess of length in respect to the lower layers - Fig. 10(b). Ultimately these structures will create a network of defects such as wrinkles and delaminations.

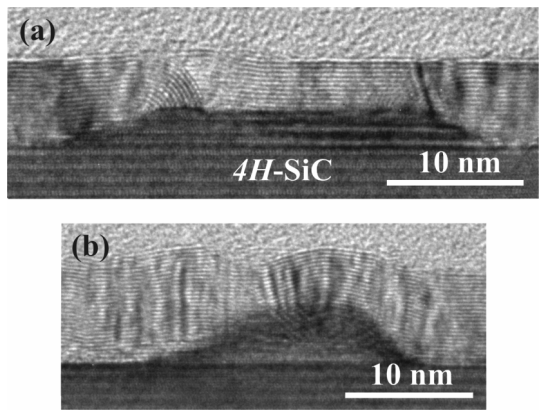

**Fig. 10.** TEM images of the disappearing SiC islands: (a) a few bending graphene multilayers, (b) coalescence of a pair of bending layers.

Images showing detachment of the graphene multilayers are presented in Fig. 11. The size of the defects is related to the morphology of the SiC surface, resulting from the etching of the substrate before annealing at high temperature.

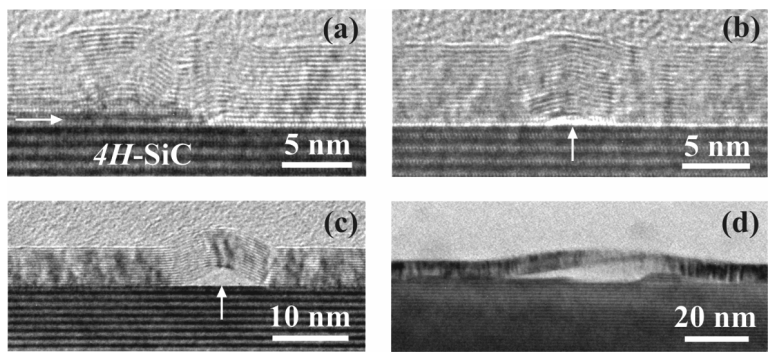

**Fig. 11.** TEM images showing detachment of the multilayer graphene: (a) as a result of disappearance of the SiC island - marked by arrow, (b) and (c) wrinkles with different sizes of channels on the flat surface, (d) wrinkle on the islands.

A different microstructure of the graphene layer synthesized on a C-terminated SiC surface was evidenced in a number of features, such as a possibility of a growth of a large number of layers, the absence of a buffer layer, a relatively large distance between the SiC topmost layer and the first carbon layer,[20] resulting in a very weak van der Waals coupling to the substrate[44] and a possible rotation of the carbon layers.[45] From the above results, it can be seen that this structure influences not only the structure but also the formation of the carbon layer, affecting the possible growth mechanism. From the above results it follows the wide space between the SiC topmost layer and the carbon layer constitutes a wide channel allowing for the effective planar diffusion of both carbon and silicon atoms. The silicon atoms escape through the channels created by the bending graphene structures. The leftover carbon atoms are incorporated into the topmost carbon layers of SiC, converting it into $sp^2$ bonded structure increasing the graphene thickness. The process proceeds without any energy barrier.

On the Si-terminated SiC surface migration of both Si and C atoms is possible at the temperature close to 1600°C. Nevertheless, creation of buffer layer, covalently bound to the silicon atoms blocks this diffusion channel, thus preventing growth of new graphene layers under the cover of topmost graphene. Therefore growth of graphene layer is terminated after creation of few carbon layers.

## V. CONCLUSIONS

The combined TEM and DFT investigations provide a scenario grasping the basic features and essential differences between the growth of graphene layers on silicon and carbon terminated SiC surfaces.

The above presented DFT data indicate the basic feature of the graphene synthesis on both surfaces of SiC crystals, at temperatures about 1600ºC used in the sublimation of a C-terminated SiC substrate to obtain carbon graphene layers. Silicon atoms easily diffuse over a Si-terminated surface, even across macroscopically large distances. Carbon atoms are less mobile but they still travel across terraces, thus allowing a growth of graphene layers which starts quickly at the steps where an escape from the lattice to the surface is facilitated. The created graphene layer is strongly attached to the SiC(0001) surface so further growth proceeds via an escape of these atoms through the lattice voids. Thus the effective growth of graphene is quickly slowed down and a few layers could be grown at the Si-terminated surface.

A different picture arises at a C-terminated SiC($000\bar{1}$) surface. Carbon adatoms are incorporated into a graphene-like lattice liberating silicon atoms from the top atomic layer. These silicon adatoms could travel far away across these terraces. The essential difference is caused by the fact that carbon layers are not covalently attached to the ($000\bar{1}$)SiC surface, thus silicon atoms diffuse in the channel underneath to find defects, such as wrinkles, which serve as an ultimate outlet of the silicon excess. The carbon atoms do not diffuse at all, they serve as graphene nucleation centers and building matter.

TEM data demonstrates a possibility of a growth of an unlimited number of graphene layers on C-terminated surfaces. The growth is supported by liberation and outdiffusion of silicon atoms from the SiC lattice. These data show that wrinkles, arising from coalescence of independently grown graphene flakes allow for an escape of silicon adatoms from the channel underneath the thick graphene layers. The data indicate a large variety of defects present in these layers, which is caused by the independent nucleation of graphene flakes under the thick graphene on the top. Thus the growth of high quality graphene at a C-terminated SiC surface is problematic.

Direct silicon sublimation from terraces needs to overcome a very high energy barrier, above 6 eV, which renders this channel ineffective. Thus another escape route has to be identified to create a full, coherent picture of the graphene growth by silicon sublimation at high temperatures.

In addition, the DFT results showing the barrierless conversion of the topmost layer of C-terminated SiC(000$\bar{1}$) surface to sp$^2$-bonded configuration, provides a nucleation center for creation of nanopipes,[46-49] during growth of silicon carbide crystals by modified Lely method in carbon-rich environment.[50]

ACKNOWLEDGMENTS

We would like to thank the Faculty of Materials Science and Engineering of Warsaw University of Technology for using the JEOL JEM 3010 transmission electron microscope. This work has been partially supported by Polish Ministry of Science and Higher Education - project 670/N-ESF-EPI/2010/0 within the EuroGRAPHENE programme "EPIGRAT" of the European Science Foundation and within the SICMAT Project financed under the European Founds for Regional Development (Contract No. UDA-POIG.01.03.01-14-155/09).